\documentstyle[12pt]{article}

\large
\newcommand{\be}{\begin{eqnarray}}
\newcommand{\ee}{\end{eqnarray}}
\begin{document}

\setlength{\baselineskip}{21pt}
\pagestyle{empty}
\vfill
\eject
\begin{flushright}
SUNY-NTG-92/40
\end{flushright}

\vskip 1.0cm
\centerline{\bf Mesonic Correlation Functions in the Random Instanton Vacuum}
\vskip 2.0 cm
\centerline{E.V. Shuryak and J.J.M.
Verbaarschot}
\vskip .2cm
\centerline{Department of Physics}
\centerline{SUNY, Stony Brook, New York 11794}
\vskip 2cm

\centerline{\bf Abstract}
\noindent
A general model-independent discussion of mesonic correlation
functions is given.
We derive new inequalities, including one stronger than Weingarten's
inequality.
Mesonic correlation functions are calculated
in the random instanton vacuum and
are compared with  phenomenological expectations and lattice results.
Both diagonal and non-diagonal correlators of all strange
and light flavored currents, as well as the most important unflavored
ones are considered.
Our results are used to extract the masses and the coupling constants
of the corresponding mesons.
Not only the qualitative behaviour is reproduced in
all channels, but in several  channels the model works
with amazing accuracy
up to distances of 1.5 $fm$.
\begin{flushleft}
SUNY-NTG-92/40\\
December 1992
\end{flushleft}
\eject
\pagestyle{plain}
\vskip 1cm
\renewcommand{\theequation}{1.\arabic{equation}}
\setcounter{equation}{0}
\centerline{\bf 1. Introduction}
\vskip .5cm
\centerline{\bf 1.1. Motivations and outline of the paper  }
\vskip 1cm

 Correlation functions are a major tool
 for understanding the structure of any type of matter, solid or liquid.
 The QCD vacuum, being a complicated ensemble of interacting quark and gluon
fields, has also been studied this way in the
framework of various theoretical
approaches based directly on quantum field theory, especially, by numerical
simulations on a lattice. However, in the past,
those studies were  mainly focused on the {\it long-distance} behaviour
of these functions, which is determined by the masses of the
lowest hadronic states.
Only  recently was it realized, that extremely
important dynamical information
is contained in the  behaviour of these functions at {\it smaller}
distances of $x\sim 0.1-1.0\, fm$.

Another explanation of why these point-to-point correlation functions are
so fundamental for QCD goes as follows.
Due to confinement, one cannot directly study quark-quark scattering, but
the same information is contained in the correlation functions.
At the same time, these functions can also
be obtained by inserting a set of physical   (asymptotic) states
and using dispersion integrals. In this way,
one can learn a lot about the space-time
structure of the effective interquark interaction.
The available
phenomenological information on correlation functions, as well as the
current theoretical situation, was
recently reviewed by one of us \cite{Shuryak_cor}.

 Although some questions related to the short-distance behaviour
of the correlators  can be traced back to 'current algebra', the Wilson
operator product expansion (OPE) and other developments in the sixties,
a systematic study of the relevant non-perturbative QCD dynamics
was initiated by the so called QCD sum rules
\cite{SVZ}. Using the OPE as a practical
tool to extract information on correlators at {\it small} distances, these
authors
have found that it works nicely in some cases, but fails
in others.

   In short, the general lesson one can draw from these studies is that
the short range behaviour of the correlation functions is
qualitatively different for different
channels. Non-perturbative effects cause deviations from the perturbative
behaviour, that differ drastically in sign and magnitude for different
channels. The resulting picture
obviously  contradicts to any simplistic model of strong interactions.
In particular,  universal color confinement
is far from being the only
non-perturbative force between quarks and anti-quarks.

   The main objective of the present work is to investigate the role of
instanton-generated effects. By now, it is well known, that at least some
of their qualitative features
(briefly reviewed in section 1.6) are in good correspondence with empirical
observations for the scalar and pseudo-scalar channels.
Moreover, numerical studies in the framework of the
'interacting instanton approximation' (IIA)  \cite{SHURYAK-1988}
show very promising results, outperforming the QCD sum rules even for
the vector and axial correlators.
In this work, we report on much more
quantitative results for many of these functions and establish more
definitely the limitations of this approach.

   Another motivation for these studies is that one can get a similar set of
results from lattice simulations. The first work in this direction
has been reported
recently   \cite{Negele_etal}. In the last section of
this work,  the comparison of  the results
obtained within these two approaches shows that they are
 surprisingly consistent with each other.

  Many details of our approach have already been
 explained in the first paper of this series
\cite{Shuryak_Ver_il1}, so we do not need to repeat them here. Let us only
remind, that
we use  the simplest possible model, the 'random instanton vacuum' (RIV),
as a kind of 'benchmark' for future investigations of instanton dynamics.
The main result of this first paper is the derivation of a
practical approximate formula for the quark propagator. More specific
 results for the {\it averaged propagator} can, roughly speaking,
 be interpreted as the
appearance of some 'quark effective mass', which happens to be about
300 $MeV$ in agreement with the old 'constituent quark model'.
 However, as we will show below, it does not mean that this model
provides a reasonable approximation to correlation functions.
The quark
propagators  contain 'hidden components'! They are not
seen in the average propagators,
but  they show up if the square
(for mesonic correlators) or the cube (for baryonic correlators) of the
propagators is   averaged over the gauge field configurations.
In other words, we have found a
strong interaction between quarks, which is not small as compared with
the 'constituent masses' and is strongly channel dependent.

  The paper is organized as follows.
In section 1.2 we remind some general 'kinematical'
properties of the correlators and introduce the currents and
the terminology used. Then, in section
1.3, we specify how the various correlators are connected to
the propagator decomposition into spin-color components. These formulae
can be used, $e.g.$, to make the simplest 'vacuum dominance
approximation', which allows us to get at least the correct sign of
the non-perturbative corrections
to various correlators induced by chiral symmetry
breaking. In this subsection we also derive some general inequalities
between hadronic correlators, including one which is stronger than Weingarten's
inequality \cite{Weingarten}.
In section 1.4, we give schematic spectral functions
for the various flavored channels. A general discussion of unflavored
correlators is given in section 1.5.
We finish the introductory part of the paper by reminding the reader of
the qualitative
features of the instanton-induced effects on correlators (see subsection 1.6).

  The main body of the paper is essentially a one-by-one discussion
of our results for the various channels, starting with a
rather complete set of 'flavored' currents (section 2)
supplemented by the most interesting 'unflavored' channels (section 3).
In all cases our results are compared with phenomenological
expectations (based on a detailed review \cite{Shuryak_cor}) or, if they are
absent, with the available predictions from QCD sum rules.
  The comparison with lattice data and a general discussion of the results
(including specific fits of masses and coupling constants for various mesons)
can be found in section 4.

\vskip 1.5cm
\centerline{\bf 1.2. General properties of the correlators    }
\vskip 0.5cm

First of all, correlators of two currents can be divided into
{\it diagonal} and {\it non-diagonal} correlators,
depending on whether they involve the
same currents or two different currents.
It is also convenient to
recognize two different types of currents,
which we call 'flavored' and 'unflavored'  currents.
The correlation function of the
flavored currents
($e.g.$ $\bar u d$ or $\bar u s$, but also $\bar u u- \bar d d$)
involves only diagrams in which both
quark and anti-quark propagate between $x$ and $y$.
The correlation function of the unflavored currents (like $\bar u u$)
also receive contributions from {\it two-loop} diagrams, in which
one quark line goes from $x$ to $x$ and another quark line goes from $y$ to
$y$.
The one-loop case, which is much simpler
($e.g.$ two-loop correlators are considered rarely in
lattice QCD calculations), will be studied first.
In line with this, we differ from the usual $SU(3)$ notation
and organization in which
the correlators are classified according to their $SU(3)-$representation.
Therefore, the  $\eta$ and $\eta'$ mesons are discussed separately
(see sections 1.5 and 3.1)  rather than together
with the $\pi$ and $K$ correlators.
Another important channel in which 'two-loop diagrams' are crucial
is the iso-scalar scalar $\sigma$ channel (see sections 1.5 and 3.2).

In this  paper we will consider the flavored currents
\setcounter{figure}{\value{equation}}
\renewcommand{\theequation}{1.1\alph{equation}}
\setcounter{equation}{0}
\be
j^S &=& \bar u d,\\
j^P &=& \bar u i\gamma_5 d,\\
j^V_\mu &=& \bar u \gamma_\mu d,\\
j^A_\mu &=& \bar u \gamma_\mu\gamma_5 d,\\
j^T_{\mu\nu} &=& \bar u \sigma_{\mu\nu} d,
\ee
\renewcommand{\theequation}{1.\arabic{equation}}
\addtocounter{figure}{1}
\setcounter{equation}{\value{figure}}

\noindent
and their counterpart with the $u-$quark interchanged by an $s-$quark. They
obey the hermiticity relation
\be
(\bar d \Gamma u)^\dagger = \bar u \Gamma d.
\ee

The study of unflavored currents will be restricted to the following
(Hermitian) currents:
\setcounter{figure}{\value{equation}}
\renewcommand{\theequation}{1.3\alph{equation}}
\setcounter{equation}{0}
\be
j^\sigma &=& \frac 1{\sqrt 2} ( \bar u u + \bar d d),\\
j^1&=& {1\over \sqrt{3}}(\bar u i\gamma_5 u+
\bar d i\gamma_5 d+\bar s i\gamma_5 s ),\\
j^8 &=& {1\over \sqrt{6}}(\bar u i\gamma_5 u+
\bar d i\gamma_5 d -2 \bar s i\gamma_5 s ),\\
j^\omega_\mu &=& \frac 1{\sqrt 2} ( \bar u\gamma_\mu u + \bar d \gamma_\mu
d),\\
j^\phi_\mu &=& \bar s\gamma_\mu s,\\
j^{f_1}_\mu &=& \bar s\gamma_\mu \gamma_5 s.
\ee
\renewcommand{\theequation}{1.\arabic{equation}}
\addtocounter{figure}{1}
\setcounter{equation}{\value{figure}}

\noindent
The two-loop diagrams contributing to the last three currents will be ignored
(Zweig's rule), and therefore they will be discussed together with
the flavored currents.

The correlators are defined by the time-ordered product
\be
\Pi(x) \equiv <0|T j_A(x) j^\dagger_B(0)|0>,
\ee
where the Dirac indices are suppressed. If we use completeness, and
PCT and
translation invariance of the vacuum state, the correlator
can be written in terms of a spectral representation as
\be
\Pi(x) = \int \frac{d^4 q}{(2\pi)^3}
(\theta(x_0) e^{-iqx} + \theta(-x_0)
e^{iqx}) \delta(q^2-\sigma^2) {\rho(\sigma^2)}d\sigma^2,
\ee
where $\rho(\sigma^2)$ is the spectral function defined by
\be
\rho(\sigma^2) &=& (2\pi)^3\sum_n \delta^4(\sigma-q_n)
<0|j_A(0)|n><n|j_B^\dagger(0)|0>,
\ee
with Dirac indices as given by the two currents.
The integral over $d^4 q$ yields a Feynman propagator of mass
$m^2 = \sigma^2$, which for spacelike distances is
given by
\be
D(\sigma,x) = \frac \sigma{4\pi^2 x}K_1(\sigma x),
\ee
where  $K_1(\sigma x)$ is a modified Bessel function. The integral in (1.5)
can be rewritten as
\be
\Pi(x) = \int_0^\infty d \sigma^2 D(\sigma, x) \rho(\sigma^2).
\ee
For spacelike $x$,
the r.h.s of (1.8) can be analytically continued to Euclidean
spacetime without hitting any singularities. This equation can thus be used to
relate the  spectral function defined in Minkowsky space-time to a correlator
in Euclidean space-time. For later use we also give its Euclidean Fourier
transform:
\be
\int d^4 x e^{iqx} \Pi(x) = \int_0^\infty ds \frac{\rho(s)}{q^2 + s}.
\ee

On the other hand, again using PCT invariance,
the spectral function can be written in terms of the
discontinuity of the correlator in Minkowsky space-time
\be
2\pi \rho(q^2) = 2{\rm Im} i\int d^4x e^{iqx} <0|T j_A(x)
j^{\dagger}_B(0)|0>.
\ee
The Euclidean Fourier transform of (1.8) with the spectral density replaced
by the integral in (1.10) yields the standard dispersion relation,
in which the correlator evaluated at Euclidean momenta
is given by an integral transform of
the imaginary part of the correlator in Minkowsky space time.

Finally, let us consider the matrix element of the current in (1.6) in more
detail. For the currents defined in (1.1) obeying the Hermiticity
relation (1.2) and the hermitian currents (1.3) the
relations
\setcounter{figure}{\value{equation}}
\renewcommand{\theequation}{1.11\alph{equation}}
\setcounter{equation}{0}
\be
<0|\bar d \Gamma u|n>^* = <0|\bar u \gamma_5 \Gamma \gamma_5 d |n>,\\
<0|\bar u \Gamma u|n>^* = <0|\bar u \gamma_5 \Gamma \gamma_5 u |n>.
\ee
\renewcommand{\theequation}{1.\arabic{equation}}
\addtocounter{figure}{1}
\setcounter{equation}{\value{figure}}

\noindent
follow from PCT invariance.
Therefore, the matrix elements of a Hermitian current are
real if the gamma matrix structure is spin flipping, $i.e.$
$\Gamma$ = $S$, $P$ or $T$, and they are imaginary if the gamma
matrix is spin nonflipping, $i.e.$, $\Gamma = V$ or $A$.

\vskip 1.5cm
\centerline{\bf 1.3. Propagator decomposition and inequalities}
\vskip 0.5cm

In the previous paper \cite{Shuryak_Ver_il1}
we already mentioned, that the
quark propagator contains both chirality-flipping and non-flipping
components. Some of them
survive the averaging over configurations, whereas others, called the
'hidden' components, are nonzero
only if a square (or cube) of the propagator is averaged. The 'hidden'
components generate the effective interaction between quarks.

Thus, the study of
mesonic or baryonic correlators actually reveals these 'hidden components'.
Generally speaking, they are by no means small as compared with
the 'visible' components. In other words, the interquark
interaction is generally of the same order of magnitude  as the
'constituent quark masses' studied in \cite{Shuryak_Ver_il1}.

In this section
we specify the  'kinematics' of this statement for various channels, and
briefly consider the relevant dynamics. The correlators, evaluated
in {\it Euclidean} space-time, can be expressed in terms of the {\it Euclidean}
propagator $S^{kl}_{\mu\nu}(x,y)$. The general structure of the correlator is
$\Pi=<{\rm Tr}(S(x,y) \Gamma S(y,x)\Gamma)>$, where the average is over all
gauge field configurations. Specifically, we discuss the following diagonal
flavored correlators:

\setcounter{figure}{\value{equation}}
\renewcommand{\theequation}{1.12\alph{equation}}
\setcounter{equation}{0}
\be
\Pi_S &=& <{\rm Tr}(S(x,y) S(y,x) )>,\\
\Pi_P &=& <{\rm Tr}(S(x,y) i\gamma_5 S(y,x) i\gamma_5)>,\\
\Pi^V_{\mu\mu} &=& <{\rm Tr}(S(x,y) \gamma_\mu S(y,x) \gamma_\mu)>, \\
\Pi^A_{\mu\mu} &=& <{\rm Tr}(S(x,y)
\gamma_\mu\gamma_5 S(y,x) \gamma_\mu\gamma_5)>,\\
\Pi^T_{\mu\nu\mu\nu} &=&<{\rm Tr}(S(x,y) \sigma_{\mu\nu} S(y,x) \sigma_{\mu\nu}
)>.
\ee
\renewcommand{\theequation}{1.\arabic{equation}}
\addtocounter{figure}{1}
\setcounter{equation}{\value{figure}}

\noindent
As we will see below, only two non-diagonal flavored correlators are
non-zero. They are given by:
\setcounter{figure}{\value{equation}}
\renewcommand{\theequation}{1.13\alph{equation}}
\setcounter{equation}{0}
\be
\Pi^{PA}_\mu &=&
   <{\rm Tr}(S(x,y)i\gamma_5 S(y,x)\gamma_\mu \gamma_5)>,\\
\Pi^{VT}_\nu &=&
   <{\rm Tr}( S(x,y)\gamma_\mu S(y,x) \sigma_{\mu\nu}>.
\ee
\renewcommand{\theequation}{1.\arabic{equation}}
\addtocounter{figure}{1}
\setcounter{equation}{\value{figure}}

The propagator can be decomposed as
\be
S^{kl}_{\mu\nu} = \sum_i a_i^{kl} \Gamma_{i,\,\mu\nu},
\ee
where the $\Gamma_i$ are a complete set of 16 Hermitian gamma-matrices
given by,
\be
\Gamma_i = {\bf 1},\,\, \gamma_5,\,\, \gamma_\mu, \, i\gamma_\mu\gamma_5,\,\,
{\rm and} \,\, i \gamma_\mu\gamma_\nu\,\,\, (\mu\ne\nu).
\ee
The coefficients $a_i$ are $SU(3)-$color matrices.
Before rewriting
the correlators in terms of the $a_i$ we point out a relation
given by Weingarten \cite{Weingarten} for the Euclidean propagator
in backward direction
\be
S(x,y)=-\gamma_5 S^\dagger(y,x)\gamma_5,
\ee
which allows us to express the diagonal correlators defined above
in terms of the positive
definite quantities
\be
Q_i = 4\sum_{kl}|a_i^{kl}|^2.
\ee
In the $V$, $A$ and $T$ case the Dirac sum over all $Q_i$ belonging to a
specific structure, is denoted by $Q_V$, $Q_A$ and $Q_T$, respectively.

The pseudoscalar ($\pi$)
correlator, $\Pi^P$, is simply given
by the sum of all coefficients squared
\be
\Pi^{P} = Q_S + Q_P + Q_V + Q_A + Q_T.
\ee
In the scalar ($\delta$) case, the correlator  $\Pi^S$ is given by the
difference of the spin non-flipping and the spin flipping components:
\be
\Pi^{S} = -Q_S - Q_P + Q_V + Q_A - Q_T.
\ee
As a result, we find Weingarten's inequality \cite{Weingarten} that the
pseudoscalar correlator should  exceed the scalar one.
{}From the experimental values of the pion and the scalar meson
masses it is clear that
this inequality should be satisfied at large distances. What is nontrivial is
that it holds  at {\it all} distances. One expects that
for $x > 0.5\, fm$ the scalar correlator is practically zero,
which implies there should be a very precise  compensation
between different components of the propagator.

The vector ($\rho$) correlator, $\Pi^V_{\mu\mu}$,
and the axial ($a_1$) correlator,
$\Pi^A_{\mu\mu}$, are given by
\setcounter{figure}{\value{equation}}
\renewcommand{\theequation}{1.20\alph{equation}}
\setcounter{equation}{0}
\be
-\Pi^{V}_{\mu\mu} &=& +4Q_S - 4Q_P +2Q_V - 2Q_A,\\
-\Pi^{A}_{\mu\mu} &=& -4Q_S + 4Q_P +2Q_V - 2Q_A.
\ee
\renewcommand{\theequation}{1.\arabic{equation}}
\addtocounter{figure}{1}
\setcounter{equation}{\value{figure}}

\noindent
At small distances both correlators are dominated by the contribution of
the free propagator and are very similar. This is no longer the case
at distances of the order of 1 $fm$ \cite{Shuryak_cor} which implies
a {\it fine tuning} of the components of the propagator. The tensor correlator,
given by
\be
\Pi^{T}_{\mu\nu\mu\nu} = -6 Q_S - 6 Q_P + 2 Q_T,
\ee
does not receive a contribution from the free propagator and is
less singular at short distances than the four other correlators.
The above relations can be inverted, which allows us to express the components
$Q_i$ in terms of the mesonic correlation functions. Only the pion
correlator always enters with a positive coefficient which implies that,
if there is only {\it one} Goldstone
boson, it necessary should be the pion. This is in agreement with the
Vafa-Witten theorem \cite{VAFA-WITTEN-1984} that vector symmetries
cannot be spontaneously broken.

Note that these relations
imply the following inequalities,
\setcounter{figure}{\value{equation}}
\renewcommand{\theequation}{1.22\alph{equation}}
\setcounter{equation}{0}
\be
\Pi^{P} &>& -\frac 38 \Pi^{V}_{\mu\mu} - \frac 18 \Pi^{A}_{\mu\mu}, \\
\Pi^{P} &>& -\frac 18 \Pi^{V}_{\mu\mu} - \frac 38 \Pi^{A}_{\mu\mu},
\ee
\renewcommand{\theequation}{1.\arabic{equation}}
\addtocounter{figure}{1}
\setcounter{equation}{\value{figure}}

\noindent
which are as strong as Weingarten's inequality. From the expressions for
the $S$, $P$ and $T$ channels we can immediately derive the inequality
\be
\Pi^P > \Pi^S + \Pi^T_{\mu\nu\mu\nu},
\ee
which is stronger than Weingarten's inequality. Because both the vector
and axial mesons couple to the tensor current (see section 1.4) it
follows that the pion mass is not only lighter than the scalar mesons but
also lighter than the vector and axial mesons, $i.e.$, $m_\pi < m_\rho$ and
$m_\pi < m_{a_1}$, which again confirms the Vafa-Witten theorem.

In the $chiral$ limit, the pion is a Goldstone mode whereas the mesons in
the other 4 channels remain massive. This requires a
precise cancellation at large separations. More
precisely, up to exponentially small corrections we have
\be
Q_S(x) = Q_P(x) = \frac 14 Q_V(x) = \frac 14 Q_A(x) = \frac 16 Q_T(x),
\quad {\rm for} \quad x\rightarrow\infty.
\ee
In other words, all components $Q_i$, in the tensor decomposition of the
propagator, converge to the same limit at large separations.

As a useful practical application of these
general relations, we derive the  corrections of the quark-condensate
to the free propagator. In the
so called 'vacuum dominance' approximation\footnote{The
reader should be warned, that 'vacuum dominance'
as introduced in \cite{SVZ} does  not lead to the corrections
we derive now, but rather to
a kind of radiative correction to it. The reason is that we consider
the OPE in space-time representation and include contributions that are
'regular' at $x=y$,  while in \cite{SVZ} the
OPE was studied at large momentum transfer, which
includes only singular terms.
See \cite{Shuryak_cor} for further discussion.},
the propagator is color diagonal and simplifies to
\be
S=S_0+ {i\over 12} |<\bar \psi \psi>|.
\ee
If the end points are taken along the $\tau-$direction, the free propagator
is given by
\be
S_0(\tau) = \frac{i\gamma_0}{2\pi^2 \tau^3}.
\ee
For the correlators in the different channels one finds
\setcounter{figure}{\value{equation}}
\renewcommand{\theequation}{1.27\alph{equation}}
\setcounter{equation}{0}
\be
\Pi^S(\tau) &=& \frac 3{\pi^4 \tau^6} - \frac 1{12}|<\bar\psi\psi>|^2,\\
\Pi^P(\tau) &=& \frac 3{\pi^4 \tau^6} + \frac 1{12}|<\bar\psi\psi>|^2,\\
-\Pi^V_{\mu\mu}(\tau) &=& \frac 6{\pi^4 \tau^6} + \frac 13|<\bar\psi\psi>|^2,\\
-\Pi^A_{\mu\mu}(\tau) &=& \frac 6{\pi^4 \tau^6} - \frac 13|<\bar\psi\psi>|^2,\\
\Pi^T_{\mu\nu\mu\nu}(\tau) &=& - \frac 12|<\bar\psi\psi>|^2.
\ee
\renewcommand{\theequation}{1.\arabic{equation}}
\addtocounter{figure}{1}
\setcounter{equation}{\value{figure}}

For completeness, let us also mention the results
for the nonzero {\it non-diagonal} correlators:
\setcounter{figure}{\value{equation}}
\renewcommand{\theequation}{1.28\alph{equation}}
\setcounter{equation}{0}
\noindent
\be
\Pi^{PA}_\mu(\tau) &=&
{2 \delta_{\mu 0}\over 2\pi^2 \tau^3}|<\bar\psi\psi>|,\\
\Pi^{VT}_{\mu\mu\nu}(\tau) &=&
{6i \delta_{\nu 0}\over 2\pi^2 \tau^3}|<\bar\psi\psi>|.
\ee
\renewcommand{\theequation}{1.\arabic{equation}}
\addtocounter{figure}{1}
\setcounter{equation}{\value{figure}}
\noindent
As we will see below, in
all channels the sign of the corrections to the free quark contribution
coincides with the
experimental trends. Thus, although the vacuum dominance approximation
does not provide a quantitative explanation of the behavior of the correlators
it certainly gives some qualitative insight into it.

Finally, we derive an inequality that can only be proved under a very plausible
but not general condition.
Let us consider the difference $Q_1 -Q_5$. It can be written as
\be
Q_1 - Q_5 = \frac 14 < {\rm Tr} (1-\gamma_5) S {\rm Tr}^* (1+\gamma_5) S >,
\ee
where the average is over all gauge field configurations. If the fermionic
modes of opposite chirality are uncorrelated, this average can be
factorized as
\be
Q_1 - Q_5 &=&\frac 14<{\rm Tr}
(1-\gamma_5) S>< {\rm Tr}^* (1+\gamma_5) S >,\nonumber\\
&=&\frac 14 <{\rm Tr} S> < {\rm Tr}^* S >,
\ee
where the second equality can  be established because
$<{\rm Tr }\gamma_5 S> = 0$ for a parity invariant vacuum state.
In terms of the mesonic correlation function the positivity of
$Q_1- Q_5$ leads to the inequality
\be
\Pi_V(x) \ge \Pi_A(x),
\ee
which is obeyed by our numerical results presented below. One also
has the inequality $\Pi_V(q) \ge \Pi_A(q)$, which  can be proved
rigorously for Euclidean momenta \cite{WITTEN-1983}.

The conditions for (1.30) are realized for a random
ensemble of instantons. In the opposite case, of an ensemble of
instanton$-$anti-instanton molecules, we have
\be
{\rm Tr} S = {\rm Tr}\gamma_5 S
\ee
before averaging over the collective coordinates. Therefore,$Q_1 = Q_5$,
and the axial and vector correlators are degenerate.

\vskip 1.5cm
\centerline{\bf 1.4. Spectral functions for flavored correlators}
\vskip 0.5cm

Next we discuss the spectral functions for the different flavored correlators.
They will be approximated by the
sum of one or two meson resonances in the form of a $\delta$
function and the contribution from the continuum. First, we consider the
$S$, $P$-case in which the coupling of the current to the resonance state is
defined by the matrix element
\be
<0|j^{S,P}(x)|p> =
\lambda^{S,P} \frac{1}{\sqrt{(2\pi)^3}} e^{-ipx}.
\ee
The total spectral function is then given by (see ref.
\cite{BJORKEN-DRELL-1965})
\be
\rho^{S, P}(s) = \lambda_{S, P}^2 \delta(s-m_{S, P}^2) +
\frac {3s}{8\pi^2} \theta(s-E_{0\,S, P}).
\ee
The continuum contribution can be obtained perturbatively. However, it
also follows from asymptotic freedom. At short distances the correlator
approaches the free correlator given by $3/\pi^4 x^6$. Alternatively,
the continuum contribution is given by the integral
\be
\int_{E_0}^\infty ds D(\sqrt{ s}, x) \rho(s).
\ee
For small $x$ the main contribution comes from large $s$. In this
region $\rho(s) \sim s$ on dimensional grounds, and by an explicit calculation
of the integral one finds $\rho(s) = (3/8\pi^2) s$.

In the vector channel the spectral function is transverse and can be written as
\be
\rho^V_{\mu\nu} =(-g_{\mu\nu} +\frac{q_\mu q_\nu}{q^2})\rho^V_T(q^2).
\ee
Its scalar part $\rho^V_T(q^2)$ is positive definite
\cite{BJORKEN-DRELL-1965}. The coupling constant
of the vector current to the $\rho$ meson is defined by
\be
<0|j_\mu^V(x)|\rho\,p> = i\lambda_\rho \frac{\epsilon_\mu}
{\sqrt{(2\pi)^3}} e^{-ipx}
\ee
where $\epsilon_\mu$ is the polarization vector of the $\rho-$meson.
Using the same approximation as in the $S$, $P$-case
the total spectral function is given by
\be
-\rho^V_{\mu\mu}(s) = 3\lambda^{2}_\rho \delta(s-m_{\rho}^2)
+ \frac{3s}{4\pi^2} \theta(s-E_{0\,\rho}).
\ee

For the axial vector channel we have the additional complication that
also the longitudinal combination contributes to the spectral function,
\be
\rho^A_{\mu\nu} =(-g_{\mu\nu} +\frac{q_\mu q_\nu}{q^2})\rho^A_T(q^2)
+ q_\mu q_\nu\rho^A_L(q^2),
\ee
where both $\rho^A_T(q^2)$ and $\rho^A_L(q^2)$ are positive definite.
The longitudinal component of the current couples to the pion.
The relevant matrix element is conventionally written as
\be
<0|j_\mu^A(x)|\pi p> = i f_\pi p_\mu
\frac 1{\sqrt{(2\pi)^3 }} e^{-ipx},
\ee
with the experimental value of $f_\pi = 133.7\pm 0.15 MeV$. The spectral
function in this case is given by
\be
-\rho^A_{\mu\mu}(s) = -f_\pi^2 m_\pi^2 \delta(s-m_\pi^2) + 3\lambda_{a_1}^2
\delta(s-m_{a_1}^2) + \frac {3s}{4\pi^2}  \theta(s-E_{0\,a_1}),
\ee
where the coupling constant
of the $a_1$ meson to the axial current is defined as
\be
<0|j_\mu^{a_1}(x)|a_1 \,p> = i\lambda_{a_1} \frac{\epsilon_\mu}
{\sqrt{(2\pi)^3}} e^{-ipx}.
\ee

Our last diagonal correlator is the tensor correlator. In this case
the contribution of the free quark propagator vanishes
which allows us
to extract the parameters of the resonances in an much 'cleaner' way.
The tensor structure of this current has two 3-vectors (like the electric
and magnetic field) with opposite $P-$parity, and thus couples to both
vector and axial mesons. One can introduce the corresponding
coupling constants as
\setcounter{figure}{\value{equation}}
\renewcommand{\theequation}{1.43\alph{equation}}
\setcounter{equation}{0}
\be
<0|j_{\mu\nu}|\rho>&=&\tilde f_\rho (\epsilon_\mu p_\nu - \epsilon_\nu p_\mu )
\frac 1{\sqrt{(2\pi)^3 }} e^{-ipx},\\
<0|j_{\mu\nu}|a_1>&=&\tilde f_{a_1} \epsilon_{\mu\nu\alpha\beta}
p_\alpha \epsilon_\beta \frac 1{\sqrt{(2\pi)^3 }} e^{-ipx},
\ee
\renewcommand{\theequation}{1.\arabic{equation}}
\addtocounter{figure}{1}
\setcounter{equation}{\value{figure}}

\noindent
resulting in the spectral function
\be
\rho^T_{\mu\nu\mu\nu}(s) = 6 \tilde f_\rho^2 m^2_\rho \delta(s-m_\rho^2) +
6 \tilde f_{a_1}^2 m^2_{a_1} \delta(s-m_{a_1}^2).
\ee

  Let us now briefly mention  the {\it non-diagonal correlators}.
We have 5 different flavored currents and therefore 10 different
pairs.  The combinations $SP$, $SA$, $PV$, $VA$ and $AT$ are
zero as a consequence of the parity invariance of the QCD action. The simplest
way to show this is to take the separation between the two
currents in the 4-direction. The desired result then follows
immediately from a parity transformation in the
path integral for the average correlator.
Two combinations, $ST$ and $PT$, are zero because an
anti-symmetric tensor cannot be constructed out of only one
vector. The $SV$ correlator is zero as a result of the conservation
of the vector current. Formally, this can be proved from
the Ward identities $<T\bar u u(x) \bar u \gamma_\mu u(0)> =0$  and
$<T\bar u u(x) \bar d \gamma_\mu d(0)> =0$.

Only the $PA$ and $VT$ non-diagonal correlators are non-zero.
In both cases the contribution of the free propagator to the corresponding
spectral function vanishes, and consequently, the continuum contribution
is absent. If 'pion dominance' is used we find
\be
\rho^{PA}_\mu(q^2) =i p_\mu \lambda f_\pi \delta(q^2 -m^2_\pi).
\ee

In the $VT$-case, the natural approximation is to saturate the intermediate
state expansion by the $\rho-$meson. The spectral function at low momenta
is then given by
\be
\rho^{VT}_{\mu\mu\nu}(q^2) =
3 i p_\nu \lambda_\rho \tilde f_\rho
\delta(q^2 -m^2_\rho).
\ee

\vskip 1.5cm
\centerline{\bf 1.5. Unflavored correlators}
\vskip 0.5cm
We consider correlators of the unflavored currents defined in (1.3).
The general structure of the correlator of
$\bar u \Gamma_A u$ and $\bar u \Gamma_B u$ is given by
\be
\Pi = <{\rm Tr} S(x,y) \Gamma_A S(y,x) \Gamma_B > +
<{\rm Tr} \Gamma_A S(x,x) \,\,{\rm Tr} \Gamma_B S(y,y)>.
\ee
The characteristic feature is the appearance of a two-loop or
annihilation diagram. The one-loop diagram is absent if the
flavor content of the two currents is different.

The vector and axial traces, $ {\rm Tr} \gamma_\mu S(x,y)$ and
${\rm Tr}\gamma_\mu \gamma_5 S(x,y)$ are not well defined for $y \rightarrow
x$.
Both traces have to be regularized by the inclusion of
the path ordered exponential $P \exp -i \int_x^y A_\mu dx_\mu$ and
the subtraction of the free propagator (which only contributes
in the vector channel). In the axial case
there will be an additional contribution due to the chiral anomaly.

However, according to Zweig's rule, all flavor changing interactions are
strongly suppressed in vector and axial channels. Therefore, we will
ignore the corresponding contributions in our calculations below and
approximate the unflavored axial and vector correlators by its one-loop
contributions only.

The situation is different in the scalar, pseudoscalar and tensor channels.
In this case the annihilation diagrams are finite. Let us consider the
correlator of the current
\be
j^{\rm  even}_u(x) = \bar u \Gamma^{\rm even} u(x),
\ee
with $\Gamma^{\rm even}$ equal to ${\bf 1}$, $i\gamma_5$ or
$\sigma_{\mu\nu}$, and a similarly defined current $j^{\rm  even}_d$.
Only the annihilation diagrams contribute:
\be
\Pi^{\rm even}_{ud} = -<{\rm Tr} S_u(x,x) \Gamma^{\rm even}\,\,
{\rm Tr}S_d(y,y) \Gamma^{\rm even}>.
\ee
According to the identity (1.16) we have
\be
{\rm Tr}\, \Gamma^{\rm even} S(x,x) = -{\rm Tr}\, \Gamma^{\rm even\,\dagger}
S(x,x),
\ee
which implies that the trace is real in the pseudoscalar case and purely
imaginary in the scalar and the tensor channel. We thus find that the
two-loop pseudo-scalar correlator is $negative$ for $x$ close to $y$
whereas the contribution from this region
to the scalar and tensor 2-loop correlator is positive.

Next, we derive the
non-diagonal spectral functions for the scalar and pseudoscalar
flavored currents.
In the scalar case $\bar u u$ and $\bar d d$ can be written in
terms of the sum and difference of $j^\sigma$ and $j^\delta$ (see eq. (1.3)).
Because
the contribution of the free quark propagator vanishes we find the
spectral function
\be
\rho^{S,\,\, {\rm 2-loop}}(s=q^2) =
(2\pi)^3\delta^4(q) <0|j^\sigma|0><0|j^\sigma|0>+
\lambda_\sigma^2 \delta(s-m_\sigma^2) - \lambda_\delta^2 \delta(s-m_\sigma^2),
\nonumber\\
\ee
which indeed gives rise to a positive correlator at short distances (the
first term in the r.h.s. has to be manipulated carefully).

The $\bar u i\gamma_5 u$ and $\bar d i\gamma_5 d$
currents can be written as a linear combination
of $j^1$, $j^8$ and $j^\pi$ which allows to express the two-loop pseudoscalar
correlator as
\be
\Pi^P_{ud} = -\frac 12 <0|Tj^{\pi^0}(x) j^{\pi^0}(y)|0>
+\frac 16 <0|Tj^8(x) j^8(y)|0>+\frac 13 <0|Tj^1(x) j^1(y)|0>.
\ee
Again, the free quark contribution to the correlator vanishes which allows
to write to corresponding spectral function as
\be
\rho^{P\,\, {\rm 2-loop}}(s) =
-\frac 12\lambda_{\pi^0}^2 \delta(s-m_{\pi^0}^2) + \frac 16 \lambda_8^2
\delta(s-m_8^2) + \frac 13 \lambda_1^2\delta(s-m_1^2).
\ee
Let us, for simplicity, consider the limit in which all quark masses are equal.
Then $m_{\pi^0} = m_8$ and $\lambda_{\pi^0} = \lambda_8$.
The $SU(3)-$singlet spectral
function, $\rho^{\eta'}$, is then given by the sum
\be
  \rho^{\eta'}(s) &=& \rho^{P\,\, {\rm 1-loop}}(s) + 3 \rho^{P\,\, {\rm
2-loop}}(s)\nonumber\\
& =& \lambda_1^2\delta(s-m_1^2) + \frac {3s}{8\pi^2} \theta(s - E_{0\,P}),
\ee
where the 1-loop pseudoscalar spectral function coincides with the flavored
pseudoscalar spectral function obtained in eq. (1.34).
Several important conclusions can be drawn from this result. First,
the continuum threshold in the $\eta'-$channel is the same
as for other pseudoscalar mesons. Second, the $\eta'-$mass is determined
by the two-loop spectral function, whereas the mass of the other pseudoscalar
mesons follows form the 1-loop spectral function. In the $\eta'-$channel the
contributions of all other pseudoscalar mesons cancel because of a detailed
compensation between 1-loop and 2-loop contribution. At relevant distances
of the order of 1 $fm$
the pion correlator is several orders of magnitudes larger than
$\eta'-$correlator, and therefore both 1-loop and 2-loop contributions
have to be calculated extremely accurately in order to determine the mass of
the $\eta'$ particle.

Finally, we show that in the quenched approximation the positivity of
the $\eta'-$correlator is violated. The simplest argument goes as follows.
Suppose, that all quark masses are equal.
Then the singlet $\eta'-$correlator is given by the sum of the one-loop
pseudoscalar correlator and $N_f$ times the two-loop correlator $\Pi_{ud}^P$.
In the quenched
approximation both contributions do not depend on $N_f$, and it is clear
that if the two contributions balance each other at one value of $N_f$, the
correlator becomes negative for $N_f+1$.

\vskip 1.5cm
\centerline{\bf 1.6. Qualitative role  of the instanton-induced interactions }
\vskip 0.5cm

   Before going into the numerical calculations of multi-instanton
effects,   let us briefly remind  the reader
of the qualitative effects of the one-instanton gauge field fluctuations.
The main physical phenomenon,
tunneling  through a barrier that separates gauge fields
of different topology, is  related to a rearrangement
of the light quark states:
some of them 'dive into the  Dirac sea' during this process, whereas others
'emerge' from it. This process   is described
by the so called {\it 't Hooft effective Lagrangian}
\cite{tHooft}
\be
L_{\rm eff}  \sim  \prod_f (\bar q_f \psi_0 )(\bar \psi_0 q_f )
\ee
where the  quark fields $q_f$ of various flavors $(f=u,\,d,\,s)$
are projected onto the so called {\it fermionic zero mode}
$\psi_0(x)$, which is the
solution of the Dirac equation $\hat D \psi_0(x)=0$ in the field of an
instanton.
The zero modes play the  role of wave functions of the states, in which
the quarks are produced or absorbed during tunneling. They
depend in a known way on collective coordinates of the instanton, its
position, size and color orientation.

An important observation is that the chirality of the zero modes
is directly related to the topological charge
of the gauge field: there is only a {\it left-handed} fermion zero mode
for an instanton and  only a {\it right-handed} fermion zero mode for
an anti-instanton. For anti-quarks, the opposite holds true.
As a result, if one ignores quark masses, it is impossible to close the
loop and return to the emitted quark because
its chirality {\it flipped} on the way.
Therefore, in the chiral limit, 'solitary' instantons are absent.
They only appear in groups of zero
total topological charge, exchanging produced  quarks among themselves.
Obviously, in order to break chiral symmetry, such  'group' has to
be infinitely large, percolating the entire space-time volume of the Universe.
The low-lying fermion modes become 'collectivized', and  occupy the entire
space-time as well, instead of being localized near a particular instanton.

At this point,  one can ask the question:
in what sense the 'one instanton approximation' can be formulated,
if instantons cannot be isolated and the low-lying
fermion modes are delocalized?
Well, provided that the instantons still form a reasonably dilute ensemble,
the delocalized modes can be written as a superposition of localized
zero modes,
 $ \psi_\lambda(x)=\sum_n C^n_\lambda \psi_{0}(x-z_n)$, where $z_n$ is
the center of the instanton. The coefficients behave as
$C^n_\lambda \sim 1/\sqrt V$, but, because $\psi(x) \sim 1/x^3$ for large
$x$, only nearby instantons contribute to $ |\psi_\lambda(x)|^2$, and for a
reasonably dilute ensemble $ |\psi_\lambda(x)|^2$ is peaked at the
center of the instantons. For the same reason,
provided the end points $x$ and $y$ (for which correlator is evaluated)
are located within the radius of a single  instanton $j$,
the quark propagator is dominated by the term
$[\psi_{0}(x-z_j)\psi^\dagger_{0}(y-z_j)]
[\sum_\lambda|C_\lambda^j|^2/(\lambda+im)]$.
The first factor, projected on quark fields, reproduces
the original 't Hooft interaction. At the same time,
the second factor,  (which for a single instanton in
the perturbative vacuum is $1/m$ (because
there is only one zero mode) is now different: it has
a non-singular chiral limit resulting from the many different modes that
contribute to the sum over $\lambda$.
Since this factor does not depend on $x-y$, it and can be ignored for
qualitative studies of instanton effects, as if only one instanton was present.

A phenomenological discussion of the instanton-induced
four-fermion interaction was first made for the pseudoscalar
channels in \cite{Geshkenbein_Ioffe}, and
similar studies have eventually led to the 'instanton liquid model'
\cite{Shuryak_1982}, \cite{Shuryak_A4}.
Below we make it even more convincing, by considering
instanton-induced effects in the scalar channels as well \cite{Shuryak_cor}.

  The specific flavor and chirality-flipping structure of the 't Hooft
interaction leads to the following important
conclusions:\\
{\it 1. The effect of the interaction is of first order
for the scalar and pseudoscalar correlators, but is absent in the vector
and axial channels.}\\
{\it 2. The sign of the corrections
is opposite for scalar and pseudoscalar channels}.\\
{\it 3. Due to its flavor structure, $\bar u u \bar d d$,
the one-instanton corrections have opposite signs for
isospin 1 and 0 correlators, which are
essentially given by $(\bar u u \pm \bar d d)^2$.}

The first point agrees with phenomenological observations \cite{Novikov_etal},
according to which much stronger
deviations from the asymptotic freedom are seen in
spin-zero channels than in spin-one channels.
The last two statements  provide a qualitative
understanding of the short distance behaviour of {\it all}
spin-zero correlators. Ignoring for
simplicity the strange quarks, we have four channels
with different parity and isospin, which we denote as
$\pi\,\,(I^P = 1^-),\,\,\eta'\,\,(I^P = 0^-),\,\, \sigma \,\,(I^P = 0^+)$
and $\delta \,\,(I^P = 1^+)$. One observes
that in the one-instanton approximation the corrections in the
$\pi$ and $\sigma$ channels have the same sign as
free propagation, which implies a larger correlator and an {\it attractive}
interaction. The opposite, with a {\it repulsive} interaction, is found in
the two other cases, the $\eta'$ and the $\delta$. This behavior of these
four channels is indeed seen in the real world, which shows
that, by studying instanton-induced interactions, we are on the right track.

  In principle, one could proceed along these lines, in which the 't Hooft
interaction is treated as an effective multi-fermion interaction that can
be taken into account systematically in perturbation theory.
However, we do not follow this approach, because
we will include {\it all} such diagrams by
an exact diagonalization of the Dirac operator in the subspace of zero-modes.

\vskip 1.5cm
\renewcommand{\theequation}{2.\arabic{equation}}
\setcounter{equation}{0}
\centerline{\bf 2. The 'flavored' mesonic channels }

In this and the following section
we present our numerical results obtained
for an ensemble of 128 instantons and 128 anti-instantons in a periodic box
of $(3.36)^3 \times 6.72\,\, fm^4$. The end points of the correlator
are taken along the longest axis, in which way most of the finite size
effects due to periodicity are suppressed even for the maximum considered
distance of 1.615 $fm$. The average correlators have been obtained by averaging
over 50 different gauge field configurations, distributed according to
the invariant measure of the integration over the collective coordinates,
and, for each configuration, by averaging over 100 different points $x$ for
a fixed value of $x - y$. The size of the instantons is taken fixed at
$\rho = 0.35 fm$.

The definition of the quark propagator in the 'random instanton
vacuum' was explained in detail in \cite{Shuryak_Ver_il1}.
Let us only comment here that for the present calculations
the quark masses have the value of 10 $MeV$ for $u$ and $d$, and 140 $MeV$
for $s$. The former is still larger than the physical mass\footnote{
Therefore, for a physical quark mass of $m_u + m_d = 11 \, MeV$,
our pion mass should be $m^{\rm calculated}_\pi = m^{\rm
experimental}_\pi (20/11)^{0.5}$.}, but it is
about as small as we can tolerate in order to be insensitive the
the artifacts \cite{Shuryak_Ver_il1}
in the spectrum of the Dirac operator at small virtualities.
The latter is not fitted to any particular quantity
but is taken as a 'common sense value'.

In order to check the volume dependence of the correlators, we also have
performed calculations with fewer instantons. A significant volume dependence
is only found for the pion and, to a much lesser extent, for the kaon. For
example, decreasing the number of instantons from 256 to 64 at constant
density, raises the pseudoscalar correlator at 1.5 $fm$ by about a factor
of 2.5, whereas 2/3 of this factor occurs by decreasing the number of
instantons from 128 to 64. Under the same conditions, the kaon correlator
changes by only 25 percent. The volume variation from decreasing the
number of instantons by a factor of two is inside the error bars for all
other mesonic correlators that have been studied.

\vskip 1.5 cm
\centerline{\bf 2.1. Two pseudoscalar channels: $\pi$ and $K$   }
\vskip 0.5cm

In this section we consider
correlators of  the flavored currents
\setcounter{figure}{\value{equation}}
\renewcommand{\theequation}{2.1\alph{equation}}
\setcounter{equation}{0}
\be
j_\pi &=& \frac 1{\sqrt 2}(\bar u i\gamma^5 u - \bar d i\gamma^5 d),\\
j_K &=& \bar u i\gamma^5 s.
\ee
\renewcommand{\theequation}{2.\arabic{equation}}
\addtocounter{figure}{1}
\setcounter{equation}{\value{figure}}
\noindent
It is obvious, that as
these pseudoscalars are the {\it lowest} excitations  of
the  QCD  vacuum, the  Goldstone modes,   they can
tell us a  great about its
{\it long-range} structure, like for example the chiral condensate.
However, unbeknownst is that  these channels
provide  very significant  information about the
{\it short-range} vacuum structure as well.

Another, but closely related,  exceptional feature of the
pseudoscalar correlators is that all components of the propagator
enter with the same positive sign (see eq. (1.18)).
Therefore, they
are the easiest to measure: no cancellations
take place. For the same reason, though, the systematic error due to the
finite volume is largest.

In Fig. 1, our results for the $\pi$ and $K$ correlators
are compared with
phenomenological expectations. As usual,
they are normalized with respect to the free quark correlator,
\be
\Pi_0(x)= {3 \over \pi^4 x^6},
\ee
which corresponds to the simple 1-loop diagram.
Therefore, the fact that our points approach 1
at small distances is
a general consequence of 'asymptotic freedom', and should be independent
of the non-perturbative vacuum fields. What is studied is
the {\it deviations} from this simple behaviour.

The solid (phenomenological)
curves have been obtained from the expression
\be
{\Pi^P(x) \over \Pi^P_0(x)}= \lambda^2_P D(m_P,x) {\pi^4 x^6 \over 3}
+ {\pi^2 x^6 \over 4} \int_{E_0}^\infty dE E^3 D(E,x)
(1+\frac{\alpha_S(E)}{\pi}),
\ee
that can be obtained from the spectral function (1.34) and the free
propagator (see \cite{Shuryak_cor} for an explanation of the perturbative
correction).
The parameters involved, namely the coupling of the pseudoscalar
current to these particles, and the 'continuum threshold' $E_0$ above which
the quarks are assumed to be produced as free ones are given by:
\be
\lambda_\pi=(480\, MeV)^2,\quad \lambda_K=1.24 \lambda_\pi,\quad
E_0=\,1.4\, GeV.
\ee
Unfortunately, the uncertainty in the absolute value of quark condensate
translates into an about 50$\%$ uncertainty of the
couplings. The uncertainty in $E_0$ is important only in a
very limited range of distances of about $0.3-0.5\, fm$,
where resonance and non-resonance contributions are comparable.
For a more detailed analysis of these parameters we refer to ref.
\cite{Shuryak_cor}.

The dashed line, shown for comparison, represents the 'vacuum dominance'
approximation (see above). Although this correction has 'good intentions',
the correct sign, it describes neither our points
nor the experimental curve\footnote{Note, that in this case
the disagreement with experiment is
not affected by the uncertainty in the value of the quark condensate,
because both the contribution of the pion
and this correction are proportional to its square.}.
More elaborate OPE formulae fail in these cases as well
(see \cite{Novikov_etal},
and \cite{Shuryak_cor}).

 The  parameters of (2.3) without the perturbative correction
can also be obtained from a fit to our
data points. The results are shown in Table 1. For the pion both
the mass and the coupling constant agree perfectly with experiment.
For the K meson the mass is right where it should be, but the
coupling constant is too small.
For a
comparison to other channels
and a general discussion of the parameters we refer to section 4.2.

\vskip 1.5cm
\centerline{\bf 2.2. Vector channels: $\rho$, $K^*$ and $\phi$ }
\vskip 0.5cm

 We use vector   currents with the same normalization
as in \cite{Shuryak_cor}\footnote{Warning: other authors, $e.g.$
\cite{SVZ}, use a different normalization.
Any definition that
includes the charge of the quarks is confusing if one compares
different channels.}:

\setcounter{figure}{\value{equation}}
\renewcommand{\theequation}{2.5\alph{equation}}
\setcounter{equation}{0}
\be
j^{\rho}_\mu &=& \frac 1{\sqrt 2}[ \bar u \gamma_\mu u - \bar d \gamma_\mu d]
                 \,\,\, {\rm or} \,\,\,\bar u \gamma_\mu d, \\
j^{\omega}_\mu &=& \frac 1{\sqrt 2}[ \bar u \gamma_\mu u +
                   \bar d \gamma_\mu d], \\
j^{\phi}_\mu &=&  \bar s \gamma_\mu s,  \\
j^{K^*}_\mu &=&\bar u \gamma_\mu  s.
\ee
\renewcommand{\theequation}{2.\arabic{equation}}
\addtocounter{figure}{1}
\setcounter{equation}{\value{figure}}
\noindent
The phenomenological expectations of the
correlators  can be obtained from the spectral function (see section 1.2).
However, in order to make contact with experiment, one uses that the spectral
density is proportional to the ratio
\be
 R_i(s) = \sigma_{e^+e^- \rightarrow i}(s)/\sigma_{e^+e^- \rightarrow \mu^+
\mu^-}(s)
\ee
of the $e^+e^-$ annihilation cross section into hadrons with quantum numbers
$i$ and the cross section
for muon pair production. Neglecting the muon mass, the latter is just
$\sigma_{e^+e^- \rightarrow \mu^+\mu^-} = (4\pi\alpha^2/3 s)$.
The trace if the spectral function in the
vector channel is related to this ratio by
\be
2\pi \rho^V_{\mu\mu}(s) = -s \frac{R^V(s)}{\pi},
\ee
which after substitution in the dispersion relation leads to
\be
<0|T j_\mu^V(x) j^{V\dagger}_\mu(0)|0>
= -\frac 1{2\pi^2}\int d s D(s^{1/2}, x)s R^V(s).
\ee
This relation can be used to extract the
'experimental definition' of the correlation
function.

   Information on strange vector  channel, marked by
$K^*(892)$, is derived
from the weak decay process $ \tau \rightarrow \nu_\tau$ + hadrons
due to
Cabbibo mixing. We also include the double-strange $\phi$ channel.
Strictly speaking, it is
related to both {\it one-loop} and {\it two-loop} diagrams, but according to
experimental observations ( the
famous 'Zweig rule'), all
flavor-changing transitions are strongly suppressed in vector channels,
and the two-loop contribution cannot be
a significant correction. All final states with a kaon pair
are included (see  details  in \cite{Shuryak_cor}) in the
phenomenological curves. Also the $\omega$ and
$\rho$ correlators only differ by  strongly suppressed flavor changing
two-loop diagrams. Within our accuracy this difference is not visible,
and therefore we only show results for the $\rho-$correlator.

Numerical results for the vector mesons
are shown by the solid lines in Fig. 2.
As usual, correlators  are  plotted in a normalized way, namely as
$ \Pi_{\mu\mu}(x)/\Pi_{\mu\mu}^0  (x)$,
where $\Pi_{\mu\mu}^0(x)= -6/\pi^4 x^6$
corresponds to the simple loop diagram, describing
free  propagation of a pair of massless quarks.

The general striking observation in all vector channels
\cite{Shuryak_C4} is that the contributions to the spectral function
of the lowest meson and other states
complement each other in such a way, that the ratio $\Pi(x)/\Pi^0(x)$
remains {\it close to one} up to the distances as large as {\it 1.5
 fm}!. This phenomenon, called  {\it
superduality},  generalizes  a well-known
'quark-meson duality' argument to much larger distances. From an experimental
point of view it means some
'fine tuning' of  the parameters of all vector states. On the other hand,
in the language of field theory it implies that in the vector channels,
for some (so far unknown!) reason\footnote{
  As was discussed briefly in section 1.6, a partial explanation
of this phenomenon is provided by the
instanton-based theory: at the one-instanton level
there is no correction to vector (and axial) correlators, as opposed to
scalar and pseudoscalar ones.},
all non-perturbative corrections to the free quark propagators
miraculously cancel each other, until the correlator has dropped
by a few orders of magnitude.

   Results of our calculations are compared with the phenomenological
expectations (solid curves in Fig.  2) obtained from the experimental
$R-$ratio as discussed above\footnote{The 1-loop radiative corrections are
included for the $\rho$ and the $a_1$ channel, but are absent in case of
the $\phi$ meson, see \cite{Shuryak_cor}.}.
First of all, {\it 'superduality'} is  qualitatively reproduced
by the calculated points.
Keeping in mind that the correlators drop
by several orders of magnitude over the range of distances under study,
even the  quantitatively agreement is actually very good.
The fact that all our curves are somewhat below the
 experimental ones came as no surprise: in the 'instanton vacuum'
under consideration there are no radiative corrections to correlators.
To first order, they are known to be
 $(1+\alpha_s(x)/\pi+...)$, which means an increase in the value of
the correlator value by
 about 10 percent. If the next terms produce  a smaller or similar
correction, the  agreement with data is actually nearly perfect.

Let us also comment that the splitting between
the  three vector channels
of flavor content $\bar u d$, $\bar u s$ and $\bar s s$
is surprisingly small up to $x = 1.5 \,fm$. As
the strange quark mass is generally
far from being negligible ($e.g.$ in the way it affects the propagators), this
observation means
that the $O(m_s)$ terms have the tendency to cancel among themselves.
The theoretical reasons for this are unclear.
At large distances, $x \sim 1.2 \, fm$, a splitting between the different
vector mesons appears, but the order is wrong: in the instanton vacuum
the $\rho$ meson is heavier than the $\phi$ meson.
  One may also look at our calculated results in a different way:
instead of comparing them
to the 'phenomenological correlator', they can be fitted directly to some
parametrization, which allows us to extract some 'hadronic parameters'.
As in the previous section, we use the standard three-parameter expression,
which now looks as follows:
\be
{\Pi_{\mu\mu}^V(x) \over \Pi^{V,0}_{\mu\mu}(x)}=
3\lambda^2 D(m,x) {\pi^4 x^6 \over 6}
+ {\pi^2 x^6 \over 4} \int_{E_0}^\infty dE E^3 D(E,x).
\ee
The coupling constant is defined as in eq. (1.37) which differs
from the definition
that is used for extracting coupling constants from the data\footnote{In
\cite{SVZ} dimensionless coupling constants are used, which
are related to ours $e.g.$ as $\lambda_\rho=\sqrt{2}m^2_\rho/g^\rho$}.
We think that in this way the comparison between the various channels
can be made most unbiasedly.

In particular,
in \cite{Shuryak_cor} we have used the matrix elements of {\it electromagnetic}
current
\be
<0|j^{\rm e.m.}_\mu |{\rm meson} > = f_{\rm meson}^{\rm e.m.}
m_{\rm meson} \epsilon_\mu
\ee
which then lead to a universal relation
containing a well measured quantity, the electromagnetic branching width:
\be
{f^{\rm e.m.}_{\rm meson}}^2={3 m_{\rm meson}
\Gamma({\rm meson}\rightarrow e^+e^-)
\over 4 \pi \alpha^2}.
\ee
 As the electromagnetic current  contains factors related  to the
electric charges of quarks
 \be j^{\rm e.m.}_\mu=
 \frac 23\bar u \gamma_\mu u-\frac 13\bar d \gamma_\mu d
-\frac 13\bar s \gamma_\mu s  + \cdots =
\frac 1{\sqrt 2}j^{\rho}_\mu + \frac 1{\sqrt 2}j^{\omega}_\mu -
\frac 13 j^\phi_\mu  +\cdots
\ee
we have to remove them which, in absence of mixing between the different
channels,  leads to the coupling constants

\setcounter{figure}{\value{equation}}
\renewcommand{\theequation}{2.2\alph{equation}}
\setcounter{equation}{0}
\be
\lambda_\rho &=& \sqrt 2 f_\rho m_\rho = (409 \pm 5\,MeV)^2, \\
\lambda_\omega &=& 3\sqrt 2 f_\omega m_\omega = (390 \pm 5\, MeV)^2, \\
\lambda_\phi &=& 3 f_\phi m_\phi = (492 \pm 15\, MeV)^2.
\ee
\renewcommand{\theequation}{2.\arabic{equation}}
\addtocounter{figure}{1}
\setcounter{equation}{\value{figure}}

\noindent
The $K^*$ coupling constant, $f_{K^*}$, is defined in the same way as
$f_\rho$ and can be obtained from the weak $\tau-$decay
(see \cite{Shuryak_cor}). For $\lambda_{K^*}$ one finds
\be
\lambda_{K^*}= \sqrt 2f_{K^*} m_{K^*} = (448 \pm 25 MeV)^2.
\ee
In table 1 these accurately known coupling constants are compared
with our calculated results.

\vskip 1.5cm
\centerline{\bf 2.3. Axial  correlators: $a_1$, $K_1$ and $f_1$ }
\vskip 0.5cm
   Currents and correlators in these channels are defined in exactly
the same way as for the
three vector channels  in the previous section with the obvious
substitution $\gamma_\mu \rightarrow \gamma_\mu\gamma_5$:
\setcounter{figure}{\value{equation}}
\renewcommand{\theequation}{2.15\alph{equation}}
\setcounter{equation}{0}
\be
j^{a_1}_\mu &=& \frac 1{\sqrt 2}[ \bar u \gamma_\mu\gamma_5 u -
              \bar d \gamma_\mu\gamma_5 d]
                 \,\,\, {\rm or} \,\,\,\bar u \gamma_\mu\gamma_5 d, \\
j^{f_1}_\mu &=&  \bar s \gamma_\mu\gamma_5 s,  \\
j^{K_1}_\mu &=& \bar u \gamma_\mu\gamma_5  s.
\ee
\renewcommand{\theequation}{2.\arabic{equation}}
\addtocounter{figure}{1}
\setcounter{equation}{\value{figure}}
\noindent
Again, we assume
that Zweig's rule is very accurate
so that annihilation diagrams can be ignored in the $\bar s s$-type
channel.

   However, as we have seen in section 1.2,
there is a significant difference: the axial correlator
correlator $\Pi_{\mu \nu}^A(q)$ has a non-zero longitudinal part,
proportional to $q_\mu q_\nu$. For zero quark masses it only contains
the contribution of massless Goldstone modes, pions $etc.$.
For non-zero quark masses,
axial currents are not conserved, and a massive pion contributes to
the spectral function.
Thus, generally speaking, in the axial cases one has two
different correlation functions which can be
split in, for example,  transverse and longitudinal parts.

   In this work we concentrate on one particular combination:
the trace $\Pi_{\mu \mu}^A$. The reason is that at this moment
we are mainly interested
in the $a_1$ meson, and not in the pion which, in this combination,
is suppressed by the small parameter $m_\pi^2$
and is unimportant at not too large distances. The details of the
extraction of the relevant $a_1$ parameters from the $\tau$ lepton decay data
can be found in \cite{Shuryak_cor}. The normalized correlator
can be represented as follows\footnote{The reader should   be
warned, that the  {\it sign} of the pion term in
this formula in \cite{Shuryak_cor} is wrong. As a result, this
correlation function changes sign and becomes negative at large distances,
where the pion term becomes dominant.}:

\be
{\Pi_{\mu\mu}^A(x) \over \Pi^{A\,0}_{\mu\mu}(x)}=
3\lambda^2_{a_1} D(m_{a_1},x) {\pi^4 x^6 \over 6}
- f^2_\pi m^2_\pi D(m_\pi,x) {\pi^4 x^6 \over 6}
+ {\pi^2 x^6 \over 4} \int_{E_0}^\infty dE E^3 D(E,x),
\ee
where the last term represents the non-resonance contribution. It is given
approximately by its asymptotic form with some threshold. Unfortunately,
the $\tau$ lepton is not heavy enough to allow its determination, and we use
two 'reasonable' values, $E_0$=1.5 $GeV$
and $E_0=$1.7 $GeV$, to show the sensitivity
of the correlator to this parameter (see two solid curves in Fig. 3).
Also in this case the 1-loop radiative corrections (not displayed in (2.16))
are included in the phenomenological curves.

  Our calculated results for the three flavored axial vector channels
are presented by the data points in Fig. 3.
As for the channels considered above, the agreement between calculations and
phenomenology is amazingly good.
All general features, including the fact that axial mesons are significantly
heavier than the corresponding
vector mesons, are obviously reproduced. Moreover, the numerical agreement
with expectations at {\it small} $x$
would be nearly perfect, if
radiative corrections would have been taken into account.

Let us also comment, that we find surprisingly
close results for $\bar u d$, $\bar u s$ and $\bar s s$ type correlators
in a wide range of distances, until approximately $x=1\,\, fm$. It means
that, as in the case of the vector channels,  in the whole region
the $O(m_s)$ terms have a tendency to cancel among themselves.

 At large distances, above 1 $fm$, the contribution of the
much lighter pseudoscalars (which we tried
to suppress, by taking the trace) still becomes dominant.
Somewhat surprisingly,
the data look good enough for the extraction of their coupling constants
and masses. Using the expression given above we have made a five
parameter fit with results shown by the dashed-dotted lines in fig. 3.

 The fitted parameters of axial mesons  and non-resonance continuum
are
listed in Table 1 and are discussed, together with others, in section 4.2.
Here we only present the results and some comments concerning the
pseudoscalars.

  For the pion the fitted values, corresponding to minimal $\chi^2$ (and to the
 curve in Fig. 3), are as follows:
\be m_\pi=252 \pm 15 MeV, \quad f_\pi= 110 MeV.
\ee
The latter value can be directly compared with
the well known experimental result of 131 $MeV$. However,
in order to
compare the calculated pion mass   to its experimental value,
one should make an adjustment for the fact that we cannot take
as small quark
masses as required by the real world.
We remind that our value for $(m_u+m_d)_{\rm calc}$=20$ MeV$,
while experimentally
$(m_u+m_d)_{\rm exp} \approx $ 11 $MeV$, and therefore
we should extrapolate the calculated pion mass to
$222 \sqrt{11/20} \,MeV = 178\pm 10 \,MeV$
which is not far from the observed value in the pseudoscalar channel
of $142\pm 14\, MeV$. In other words, the model does
reproduce the right magnitude of the
mass-independent coefficient  $m^2_\pi/(m_u+m_d)$ for small enough quark
masses.

  The corresponding fitted values of the parameters in (2.16) for the kaon are:
\be
m_K=464 \pm 15 MeV, \quad f_K= 100 MeV.
\ee
The mass is right on the experimental value, but the decay constant
is about a factor 2 smaller than needed.

For the 'purely strange'
pseudoscalar $\eta_s$ we find from the fit:
\be
m_{\eta_s} = 553\pm 15 MeV, \quad f_{\eta_s}= 90 MeV.
\ee
 As this combination is actually some
mixture of $\eta$ and $\eta'$, with the two-loop effects so far ignored
(see below),
we do not think these results have a direct physical meaning, but
we present them for completeness.

\vskip 1.5cm
\centerline{\bf 2.4. Two 'flavored' scalars  }
\vskip 0.5cm

At this point we would like to make some remarks on the accuracy of our
calculations. All calculations have been performed in the quenched
approximation where the small eigenvalues are not suppressed by the
fermion determinant. In the one-instanton approximation the vector
and axial correlators have nevertheless a finite chiral limit, and
we expect that the enhancement in the spectrum for small
virtualities \cite{Shuryak_Ver_il1}
does not significantly alter the correlator.
However, in the case
of the scalar, pseudoscalar and tensor correlators, the chiral limit
cannot be taken in the one-instanton approximation, and therefore
we should not be surprised if we are unable to reproduce the experimental
results with the physical values of the light quark masses.
Because the pseudoscalar correlator is the sum of all 'hidden components'
the relative error is still small, and, in section 2.1, we have found that
the phenomenological curves can be obtained with somewhat larger quark masses.
On the other hand, in the
case of the flavored scalar correlators, the first-order instanton-induced
interaction is 'repulsive at small distances (see section 1.6).
At larger distances, other corrections ($e.g.$  multi-instanton ones)
become important, but these correlation
functions are small as compared to those considered previously.
At a distance of the order of 1 $fm$ a cancellation of a couple of orders
of magnitude has to take place
between the spin flipping and the spin non-flipping components of
the propagator. We necessary
will find a large systematic error not only because of the anomaly in the
spectrum at small virtualities, but also because we use approximate
propagators. In the tensor channel the
instanton induced interactions are attractive. Consequently, we have
much less cancellations and much more accurate results can be obtained.
The non-diagonal
correlators are less singular in the chiral limit (see, for example, the
vacuum dominance results), and also in this case we expect
smaller errors than in the scalar case.

After these warnings, we are ready to confront our results for the scalar
correlator.
Our results for the $\bar u d$ and $\bar u s$ channels are shown by the
data points in Fig. 4.  One observes that our data decrease more rapidly
than in any other channel considered above. In fact, our data even 'overshoot'
this trend.  The 'random instanton vacuum' produces a
too strong repulsion in these
channels: the correlators
become negative (since we do not trust the results for the correlator in this
region we did not even plot those points).
This violates the positivity of the scalar correlator
and it points to a defect of the model. Indeed, as we have stressed above,
all calculations have
been performed in the quenched approximation.

Apart from the $a_0(980)$ particle
(which is believed to be mainly an $\eta \pi$ or
a $\bar K K$ system with flavor composition
essentially given by $\bar u d \bar s s$), {\it no} flavored
scalar mesons are listed by Particle Data Group, so, probably,
they do not exist at all!
However, continuum (multi-meson)
states should of course exist, and, for comparison, we have plotted
some 'expectations' in Fig. 4. We have assumed the absence of
resonances and have pushed continuum threshold  up to $E_0= 2$ GeV, which we
consider as a kind of its upper limit. Therefore,
the solid  curve
represents  {\it the most rapidly decreasing} correlator, on the edge of what
seems reasonable.

\vskip 1.5cm
\centerline{\bf 2.5. Tensor and non-diagonal correlators }
\vskip 0.5cm

  This section deals with a set of correlators which
 are only rarely discussed in literature, $e.g.$ they are not
mentioned in  reviews on the subject. However, as we are going to show
shortly, they are actually quite useful,
because all of them can be described very well by the dominance of
one intermediate state. Therefore, they are much more suitable
for the determination of hadronic masses than
all standard correlators.

  The first example is the correlator of two  'tensor currents'
$j_{\mu\nu}=\bar u \sigma_{\mu\nu} d$ (and
also its strange analog, $d\rightarrow s$).
As the two indices are anti-symmetric in this case, the intermediate
physical states are not the usual tensor mesons, of course, but rather some
specific polarization states of vector and axial mesons.

Contrary to all other diagonal correlators
considered above, these tensor correlators do not have
a free-quark contribution at small distances: the corresponding
loop diagram vanishes kinematically because of the
Dirac trace alone. (Therefore,  one hopes to obtain
the contribution of
resonances in a much 'cleaner' way.)
The coupling
constants were defined in (1.43), and, in this case the normalized correlator
simply follows from eq. (1.44) with result
\be
\frac{\Pi^T_{\mu\nu\mu\nu}(x)}{\Pi_0^P(x)}=
2\pi^4 x^6 \tilde f_\rho^2 m^2_\rho D(m_\rho,x)+
2 \pi^4 x^6 \tilde f_{a_1}^2 m^2_{a_1} D(m_{a_1},x).
\ee

  Our data are shown in Fig. 5 together with an
approximate fit, corresponding to
 $\tilde f_\rho=100\, MeV$, $\tilde f_{a_1}= 0.0 MeV$
 and experimental $\rho$ and $a_1$ masses. Thus, this correlator is
completely dominated
by a single rho meson pole.

   Our next topic is the non-diagonal
vector-tensor correlator.
This is the only non-diagonal correlator
that was  considered in the literature in connection
with QCD  sum rules
\cite{Balitsky_Yung,Balitsky_Kolesnichenko_Yung,Ioffe_Smilga}.
It is related
to the baryon magnetic moments and other
electromagnetic properties of hadrons. In particular, the
susceptibility of the quark condensate $\chi(q^2)$ is defined by
the small-$q$ limit of the integral
\be
\int d^4x e^{iqx} \Pi^{VT}_\mu = 3 i \chi(q^2) <\bar q q> q_\mu.
\ee
The correlator can also be expressed in terms of the spectral function.
When substitute the $\rho-$meson dominance formula (see eq. (1.46)) in eq.
(1.9) we find the sum-rule
\be
\chi(q^2) = \frac{\lambda_\rho \tilde f_\rho}{<\bar q q> (q^2+ m_\rho^2)}.
\ee
For phenomenological values of $m_\rho$, $\lambda_\rho$ and $\lambda_\rho$
and our fitted value of $\tilde f_\rho$ we find $\chi(0) = -2.0 \, GeV^{-2}$
which differs from the corresponding value $\chi(0) = -3.3 \, GeV^{-2}$
obtained in \cite{Balitsky_Kolesnichenko_Yung}. This deviation
originates in the coupling constant $\tilde f_\rho$ which was estimated with
the help of vacuum dominance in \cite{Balitsky_Kolesnichenko_Yung},
$i.e.$, $\lambda_\rho \tilde f_\rho = 2 <\bar q q>$, resulting in
$\tilde f_\rho = 165\, MeV$ as opposed to our value of 100 $MeV$.
A more elaborate QCD sum rule calculation taking into account two additional
$\rho-$meson resonances
\cite{Balitsky_Kolesnichenko_Yung} gives a slightly larger value
of $\chi_{\rm e.m.}=-4.4 GeV^{-2}$ (at the normalization point $\mu_0=1$ GeV).

More explicitly, they used the spectral function
\be
\rho_\mu(q^2) = 3 i q_\mu [c_\rho \delta(q^2-m_\rho^2) + c_{\rho'}
\delta(q^2-m_{\rho'}^2)  + c_{\rho''} \delta(q^2-{m_{\rho''}}^2)],
\ee
where the three terms contain the contributions
from the three lowest $\rho$ resonances,
with masses of 780, 1250 and 1600 $MeV$,
respectively. The constants are related to the coupling constants and
are equal to $c_\rho = 2.9 |<\bar q q>|$, $c_{\rho'} = -1.1|<\bar q q>|$
and $c_{\rho''} = 0.6 |<\bar q q>|$. The ratio of the corresponding $VT$
correlator to the free pseudoscalar correlator $\Pi^P_0$ given by
(for $x_0 > 0$)
\be
\frac{\Pi_{\mu\mu\nu}^{VT}(x)}{\Pi_0^P(x)} = -\pi^4 x^6 [
c_\rho \frac {d D(m_\rho,x)}{d x_\nu}+
c_{\rho'} \frac {d D(m_{\rho'},x)}{d x_\nu}+
c_{\rho''} \frac {d D(m_{\rho''},x)}{d x_\nu}]
\ee
is represented
by the solid curve in Fig. 6a.  Let us remind, there are no free parameters
and that we have an absolute normalization.
In the same figure, these predictions
are compared with our results. One should note
that although the two  calculations are completely different in nature
the agreement is quite reasonable.

At small distances the agreement is exceptionally good.
Of course, this is not accidental:
QCD sum rules are based on the OPE, and
should do better in this region.
In fact, the OPE dictates that the short distance behaviour of $\Pi^{VT}$
is given by the vacuum dominance formula (1.28b).
Therefore, the  excellent agreement at very small distances
is essentially do to the fact that we reproduce the right
magnitude of the quark condensate.

A natural generalization of
electromagnetic susceptibility of the quark condensate to the case
of a {\it weak external axial
current} \cite{Belyaev_Kogan} leads to another non-diagonal
correlator: the
pseudoscalar-axial correlator. It can be  used to evaluate the
axial coupling constants of various baryons (see also \cite{axial_field}), but
presently we are interested in the correlator by itself.

   One can certainly argue that in this case one  knows at least its
long-range part, which follows from the pion intermediate state.
Moreover, there are no unknown constants
because the coupling of the pion to axial current is nothing else
but the famous pion decay constant (see eq. (1.38)).
Using 'pion dominance' for this non-diagonal correlator
we find the expression
\be
\frac{\Pi^{PA}_\mu(x)} {\Pi_0^P(x)}= \frac{\pi^4 x^6}{3}\left [
-\lambda_\pi f_\pi  \frac{d D(m_{\pi},x)}{d x_\mu}\theta(x_0)
+\lambda_\pi f_\pi  \frac{d D(m_{\pi},x)}{d x_\mu}\theta(-x_0)\right ].
\ee
In Fig. 6b, we show the numerical data together with the above
formula with a fitted value of $f_{\pi}$ which shows agreement at all
distances.
If one takes $m_\pi = 191 \,MeV$ (bcause $m_u + m_d = 20\, MeV$
and $\lambda_\pi=(510 MeV)^2$,
as fitted from pseudoscalar correlator, the result for pion decay constant
reads
\be f_\pi = 110 \pm 17 MeV \ee
which is the same value as obtained from the axial channel.

   The final remarks of this chapter deal with scalar-vector correlator.
Our calculations show results that are consistent with zero
at all distances.
The difference between $SV$ and $PA$ correlators is
seen from the following argument. Suppose that some intermediate scalar (or
pseudoscalar) particle contributes to it, then we have a contribution
of the type (2.25). Let us now take another derivative $\partial_\mu$ of the
correlator: when it acts on the propagator
it produces the mass squared of the intermediate
state. But, in the chiral limit, both the vector and axial
 currents are conserved: the result should be zero.
The dilemma is then as follows: (i) the coupling constant is zero, or (ii)
the particle mass is zero. In the $PA$ channel the second alternative
is realized, whereas in the
$SV$ case only the first one is possible.

  Therefore, a zero $SV$ correlator implies that
the vector current is conserved in
our calculations (see section (1.4)).
Of course, this holds true rigorously if a complete set of states is
used in the propagator.
In practice, we work with an incomplete set of states,
so the vanishing results are actually a
non-trivial test of our calculations.

\vskip 1.5cm
\renewcommand{\theequation}{3.\arabic{equation}}
\setcounter{equation}{0}
\centerline{\bf 3. Correlators of some 'unflavored' currents  }
\centerline{\bf 3.1. Pseudoscalar channels  $\eta$ and $\eta'$  }
\vskip 0.5cm

   As explained above,
both {\it one-loop} diagrams and 'two-loop' (or annihilation)
diagrams should be included in the 'unflavored' channels (see section 1.5).
It implies that  the
propagators should be defined at {\it zero} separations between
its end points,
or that the regular part of the propagator should be separated from the
singular part. Generally speaking, this is non-trivial:
for example, the regular part of the axial current
should reproduce correctly the axial anomaly.
Fortunately, significant simplifications take place
for pseudoscalar currents because the  Dirac trace kills all
chirality non-flipping terms in the propagator.
In practice, the 'double-loop' diagram was calculated from
the zero-mode part of the propagator only, which is non-singular by
construction.

  Let us first explain the definition of the correlators.
For two massless  flavors, we can distinguish
 two different components
\setcounter{figure}{\value{equation}}
\renewcommand{\theequation}{3.1\alph{equation}}
\setcounter{equation}{0}
\be A(x)&=&<T\bar u i\gamma_5 d(x)\bar d i\gamma_5 u(0)> ,\\
 B(x)&=&<T\bar u i\gamma_5 u(x)\bar d i\gamma_5 d(0)>,
\ee
\renewcommand{\theequation}{3.\arabic{equation}}
\addtocounter{figure}{1}
\setcounter{equation}{\value{figure}}

\noindent
related to one and two-loop diagrams respectively.
Inclusion of a massive $s$ quark
brings in three more functions,
\setcounter{figure}{\value{equation}}
\renewcommand{\theequation}{3.2\alph{equation}}
\setcounter{equation}{0}
\be C(x)&=&<T\bar u i\gamma_5 u(x)\bar s i\gamma_5 s(0)>,\\
D(x)&=&<T\bar s i\gamma_5 s(x)\bar s i\gamma_5 s(0)>,\\
E(x)&=&<T\bar u i\gamma_5 s(x)\bar s i\gamma_5 u(0)>,
\ee
\renewcommand{\theequation}{3.\arabic{equation}}
\addtocounter{figure}{1}
\setcounter{equation}{\value{figure}}

\noindent
with $D(x)$ possessing both  contributions, and $C(x)$, $E(x)$ only one of
them.
All 5 functions can be obtained from correlators in the four light mesonic
channels
 $\pi$, $K$, $\eta$ and $\eta'$, plus one non-diagonal
correlator $\eta-\eta'$.

  Although we do not really use the
strange quark mass as a small parameter, we
still use the $SU(3)$ singlet and octet currents defined in eq. (1.3).
In this section we focus on the following 3 correlators
\setcounter{figure}{\value{equation}}
\renewcommand{\theequation}{3.3\alph{equation}}
\setcounter{equation}{0}
\be
\Pi^{11}(x)&=&<Tj_1(x)j_1(0)> = {1\over 3}(2A+4B+4C+D),\\
\Pi^{88}(x)&=&<Tj_8(x)j_8(0)> = {1\over 3}(A+2B-4C+2D),\\
\Pi^{18}(x)&=&-<Tj_1(x)j_8(0)> ={\sqrt{2}\over 3}(-A-2B+C+D).
\ee
\renewcommand{\theequation}{3.\arabic{equation}}
\addtocounter{figure}{1}
\setcounter{equation}{\value{figure}}

\noindent
  Note, that in the $SU(3)$ symmetric case ($m_s = m_u = m_d$), $B=C$
and $D=A+B$.
Therefore $\Pi^{88}=A$ and $\Pi^{18}=0$, as they should:
the $\eta$ and $\pi^+$ channels are the same,
 and there is no mixing.
However, due to the chiral anomaly, $\Pi^{11}=A+3B \neq \Pi^{88}$.

In Fig. 7, our results are compared with the phenomenological expectations
\cite{Shuryak_cor}. The diagonal $\eta$ and $\eta'$ correlators are shown
in Fig. 7a. The obvious
observation is that while the model is doing an excellent job in
reproducing the $\eta$ curve, it certainly 'overshoots' in the $\eta'$
case: the positivity condition of the correlator is violated. In the $\eta$
case the two-loop contributions nearly cancel among themselves
(exact in the chiral limit), whereas
all one-loop contributions add up with the same sign. The $\eta'-$meson
remains massive in the chiral limit which requires a delicate balance
between the positive one-loop and negative two-loop contributions
(see section 1.5).
The latter are closely related to the correlations of the
topological charge density, which are not implemented properly in the
'random model'.

  In Fig. 7b we show our results for
$\eta-\eta'$ mixed correlator. Let us remind that idea of
mixing was introduced in
the days of non-relativistic quark model. One can imagine that the
singlet and octet states $\eta_1$ and $\eta_8$,
produced by the singlet and octet currents at low energies,
are a mixture of the physical $\eta$ and $\eta'$ states.
The non-diagonal correlator should look as
\be
\Pi^{18}= \lambda_\eta \lambda_{\eta'} \cos\theta \sin\theta
[D(m_\eta,x)-D(m_{\eta'},x)] + ({\rm mixing\,\, in\,\, the\,\,continuum}).
\ee
The curve
shown for comparison uses $\lambda_\eta=1.3  \lambda_{\pi}$,
$\lambda_{\eta'}=0.7 \lambda_\pi$ and
a angle of $\theta=20^o$. The sign and qualitative behaviour is
right, but the angle (or couplings?) seems to be larger than expected.
However, much more work is needed in order to understand this mixing
phenomenon properly.

\vskip 1.5cm
\centerline{\bf 3.2. The isoscalar scalar (or $\sigma$) channel }
\vskip 0.5cm

This channel
 is different from all others considered above, because
 the corresponding current
\be j^\sigma={1\over \sqrt 2} (\bar u u + \bar d d )\ee
has a non-zero vacuum expectation value. As a result, the
correlator in question has
a non-zero {\it disconnected part}
\be
\Pi^\sigma(x) \rightarrow 2 |<\bar q q>|^2
\ee
that dominates at large distances.

  Therefore, our data are plotted  in two different ways.
In Fig. 8a we show the measured signal, normalized,
 as usual, to the perturbative
correlator.  The solid line shows the phenomenologically
expected  contribution of $2 \pi^4|<\bar \psi \psi >|^2 x^6/3$
(for  $<\bar \psi \psi >=-(240 MeV)^3$).
One observes that the correlator exceeds the perturbative
correlator by a big factor, and that
our points are somewhat above the phenomenologically expected (solid) curve.
The reason is that our model produces a
 slightly larger value of the condensate, namely,
$ |<\bar \psi \psi >|=2.22 fm^{-3}=(257 MeV)^3$, which perfectly agrees
with the large distance behavior of our points.

   However, the disconnected signal is not very interesting
by itself. The physical excited states are related to
the 'connected' part of the correlator. Unfortunately, as shown
in Fig. 8b, the accuracy at large distances is greatly reduced
{\it after} the subtraction of the
disconnected part. It is important to note, that the connected part
of the correlator is still quite large.
This means that the
effective interquark interactions are indeed {\it strongly attractive}.
Of course, this is hardly surprising
because in the one-instanton approximation
this should be the case (see section 1.6).

As can be seen from  the dashed lines in Fig. 8b,  the connected
part can be very well fitted by our standard three-parameter
fit, resonance plus continuum as in (1.34).  The following
values of the parameters were obtained:
\be m_\sigma=543\, MeV, \,\,\, \lambda_\sigma=
(500 \, MeV)^2,\,\,\,E_0 = 1160\, MeV
\ee

  Those parameters  show that {\it the 'instanton vacuum'
contains the 'sigma meson'}\footnote{
Of course, on the basis of these data
one cannot answer the old question, whether the 'sigma-meson'
is a resonance or just a well-correlated $\pi\pi$ pair, but one definitely
sees that in order to reproduce our results for the correlator one should have
a peak in spectral density around 600 MeV.} the
famous enhancement in the $\pi\pi$ cross section which plays a prominent
role in nuclear physics and is the basis of
the 'sigma model' of Gell-Mann and Levy.

 Moreover, the particular parameter determined best of all
is the sigma coupling constant. It is interesting to note, that it
  happens to be {\it the largest} of all
mesons. It implies a very compact sigma particle, obviously due to
 the very strong attractive interaction in this channel.

\vskip 1.5cm
\centerline{\bf 4. Discussion }
\centerline{\bf 4.1. Comparison with other works }
\vskip 0.5cm
   In the discussion above we have compared
our results for correlation functions with phenomenology. Now, we are
going to compare them with other theoretical results.
There generally are three different categories:
(i) the operator product expansion (OPE) in the context of QCD sum rules
with many parameters (or 'condensates');
(ii) other instanton-based calculations; and (iii) lattice QCD calculations.

   We will not discuss the OPE-based expressions, partly because it was
done in detail in \cite{Shuryak_cor}, and partly because we already
discussed its simplest
version (based on the quark condensate contribution) in section 1.4.
For the tensor and non-diagonal
correlators we have given a more elaborate discussion because
no other sources of information are available.

   The instanton-induced effects were originally discussed
in \cite{Geshkenbein_Ioffe,Shuryak_1982} to the
{\it first-order in  instanton density}.
In section 1.6 we have given a qualitative discussion which is
even somewhat wider than those works, because it also includes
the scalar channels.

Further developments include the analytic attempts to account for
the {\it multi-instanton} effects by means of summing a subset of diagrams
\cite{MCDOUGALL-1983,CARLITZ-1984,Diakonov_Petrov_mesons},
\cite{CARNEIRO-MCDOUGALL-1984}.
The first numerical study of correlators in the interacting instanton vacuum
\cite{SHURYAK-1989D} showed surprisingly good results.
We do not discuss them here for two reasons. (i)
Within their much larger statistical and systematical uncertainty
the old calculations agree with the present
ones. (ii) A much more accurate and detailed
study of the 'interacting instantons' vacuum in under way and
will be discussed elsewhere \cite{Shuryak_Verbaarschot_IIA}.

In the last part of this section we compare
our results with the first lattice calculations
  of {\it point-to-point} correlation functions, which were
recently reported in \cite{Negele_etal}. Results for four non-strange
channels, shown in Fig. 9, show a qualitative
agreement in all cases. In the $\pi$ and $\rho$ channels one even
finds, somewhat  curiously, that the phenomenological curve is in between
the two calculations.
The axial channel agrees very well, while in the isovector scalar ($\delta$)
case our results definitely 'overshoot' the repulsion (see discussion above).

As one compares these two sets of data, resulting from
completely different theoretical approaches,
one should keep in mind that the plotted
curves do not really represent the measured signal:  for convenience
of representation the correlators are
normalized with respect to the perturbative correlators.
Actually  the measured correlators change by several orders
of magnitude, so their deviation at the largest distances
by, say, a factor of
two is in fact a surprisingly good  agreement.
As we will show in our next paper of this series
\cite{SHURYAK-VERBAARSCHOT-1992E}, this agreement is
observed in the baryonic channels as well.

   The reason for the intriguing agreement between the two set of
correlators can and should be investigated further.
The most obvious way to proceed is,
of course, to 'cool' the lattice configurations until one is left with
mostly instantons, and then to redo the measurements of the correlators.
If the resulting
correlators happen to be about the same, our main point
{\it the dominant role of the instanton-based
 quark interactions in the QCD vacuum}, would finally be
 demonstrated explicitly.

\vskip 1.5cm
\centerline{\bf 4.2. The fitted masses and coupling constants }
\vskip 0.5cm

   In our discussion above we have concentrated on {\it correlation functions}
as such and compared results directly with phenomenological curves
\cite{Shuryak_cor} based on the integrated physical spectral density
or lattice data \cite{Negele_etal}. We do believe it is the best way
to confront the model under consideration with reality.

The extraction of hadronic parameters from correlation functions by fitting
is easily beset with systematic errors, for example, because they
are very sensitive to the parametrization of the fitting function.
We have performed two types of fits: (i) using the data points with
statistical errors only
(for results given in Table 1),
and (ii) with statistical errors provided they are
larger than 5 percent, otherwise they are put to be  5 percent. The difference
is related to a different weight assigned to the small $x$ region,
for which the statistical
errors are very small. Generally, the difference in the fitted values
is of the order of  10 percent, except in special
 cases like the
 $\rho$ mesons, where the $\chi^2$ has a complicated valley
in the parameter space,
so that small changes in the calculated points may drastically change the
fitted parameters..
Lattice data typically resolve such problem by fitting only the long-distance
part of the correlators, ignoring the small $x$ behaviour.

  Having said this, we still would like to report the results of such fits.
All points, from $x=0$ until the largest $x$,
are fitted with 'pole plus continuum'
expressions (two poles for axials), with only statistical errors included.
A subset of results
for diagonal 'flavored' correlators is listed in Table 1.
Let us make the following comments.

\begin{itemize}
\item[--] The masses of the lowest resonances in these channels,
ranging from light pions to heavy axials,
are very  well reproduced  by the model. The largest deviation,
for the $\rho$ meson, is about 15 percent.

\item[--] The coupling constants $\lambda$ are somewhat
smaller than the experimental values. It is probably due to the fact
that the model does not have confinement, which effectively
cuts off the tails of the wave functions and makes the  particle more
compact. The largest deviation is observed for the $\phi$ meson which is
However,
with our definition\footnote{Let us
remind the reader, that it differs
from standard definition in literature. This allows us to compare
different channels in a more straightforward way.} of $\lambda$,
the model correctly reproduces the following
natural trend: the coupling constants decrease from the
'most attractive' channels ($\sigma,\,\, \pi$) to the
'least attractive' ones (vectors and axials). In the
most 'repulsive' channels (flavored scalars) the resonances
are absent altogether.

\item[--]  In contrast to the parameters
of the lowest excitations, the
position of 'continuum threshold' is surprisingly channel independent.
Unfortunately, we do not have much
phenomenological information about these
important parameters (except for vector channels). However, if we assume that
the threshold coincides with the position of
the {\it next to lowest} state, the primed resonance,
these numbers can be compared with the experimentally known
 'primed' resonances. Indeed, in all known cases they are in
the region around 1200 MeV, as the fitted thresholds indicate. Thus, one may
speculate that the RIV model does
 predict the  correct position of the 'continuum threshold' as well.
\end{itemize}

\vskip 1.5cm
\centerline{\bf 4.3. Summary }
\vskip 0.5cm

   We have reported  a detailed calculation of a wide set of
mesonic correlation functions in
the  'random instanton vacuum'. The original idea was to use
a crude
model as a benchmark for further studies of {\it
interacting instantons}, with the
parameters (the instanton density and a typical radius) as
 suggested a decade ago \cite{Shuryak_1982}, without any
further adjustments.

However, the results obtained
show that this model  works amazingly well, in particular for
the pseudoscalar correlators. Asymptotic freedom is recovered at
small $x$, and at intermediated distances the correlation function
can be parametrized by a spectral function with one pole (two for
axials) and a continuum contribution.
At the moment it is not
clear why the phenomenological correlators are reproduced so well in spite
of the apparent
absence of such usual ingredients as confining forces and perturbative
effects.

  In fact,  not only
the  qualitative behaviour of all correlators at
distances as large as 2 $fm$ is reproduced, but in
a few cases, when a quantitative analysis is possible, we found agreement
with the physical parameters at the 10-20 percent level.
Thus, it is certainly a useful tool, so wider use of the model
seems to be justified.

  At the same time, in two channels with a strong repulsive interaction,
the $\eta'$ and the $\delta$, the results are not satisfactory: the model
correctly predicts such repulsion but
overestimates its magnitude.
 We are looking forward to more realistic
calculations with 'interacting instantons'  and hope to improve on
this situation.

\vskip 1cm
\centerline{\bf 5. Acknowledgements }
Numerical studies reported in this work
 were mainly performed on a CRAY machine at NERSC, which is
greatly acknowledged. This work is partly supported by the DOE grant.
We would like to thank J. Negele for making
results of his lattice studies available to
us prior to publication.

\newpage
\begin{tabular}{||l|l|l|l|l||} \hline
 channel $j^P$ & $m_{\rm res}\, [MeV]$ & $\sqrt{\lambda}\, [MeV]$
 &  $E_0$ [MeV] &   comment    \\ \hline
$\pi, 0^-$  & 142 $\pm$ 14  & 510$\pm$20 &  1360 $\pm$ 100 & this work \\
  &  138  & 480 &   & phenomenology \\ \hline
$K, 0^-$  &  482 $\pm$ 12 & 467$\pm$20 &  1350$\pm$50 & this work \\
  & 495  &  & 595  & phenomenology \\ \hline
$\eta, 0^-$  & 500  & 462 &  1250 & this work \\
  & 548
.8  &  &   & phenomenology \\ \hline

$\sigma, 0^+$ &543 & 500 & 1160 & this work \\ \hline

$\rho, 1^-$  & 950$\pm$100  & 390$\pm$20 &  1500$\pm$100 & this work \\
  & 780  & 409$\pm$ 5&   & phenomenology \\ \hline
$K^*, 1^-$  & 860$\pm$15  & 341$\pm$20 &  1300$\pm$50 & this work \\
  & 892  & 448$\pm$25 &   & phenomenology \\ \hline
$\phi, 1^-$  & 850$\pm$50  & 280$\pm$20 &  1000$\pm$40 & this work \\
  &  1020  & 492$\pm$15  &   & phenomenology \\ \hline
$a_1, 1^+$  & 1132$\pm$50  & 305$\pm$20 &  1100$\pm$50 & this work \\
  &  1260  & 400 &   & phenomenology \\ \hline
$K_1, 1^+$  & 1170$\pm$50  & 302$\pm$20 &  1120$\pm$50 & this work \\
  & 1270 &  &   & phenomenology \\ \hline
$f_1, 1^+$  & 1210$\pm$50  & 293$\pm$20 &  1200$\pm$50 & this work \\
  & 1285 &  &   & phenomenology \\ \hline
\end{tabular}
\vskip 1cm
\centerline{\bf Table 1}
\vskip 1cm
The mass $m_{\rm res}$, the coupling constant $\lambda$ and the threshold
energy $E_0$ of some mesons in absolute units fitted according
to the three-parameter formula for the spectral function. The pion mass
is adjusted to the physical quark masses under the assumption that
it is proportional to $(m_u+m_d)^{1/2}$.
Whenever available the parameters are
compared with  experimental numbers.

\addtocounter{page}{+2}
\newpage
\centerline{\bf Figure Captions }
\vskip 0.5 cm
\noindent
Fig. 1. Pseudoscalar correlation
functions with quantum numbers of the pion and the kaon normalized to free
massless quark correlator in the same channel
plotted versus the distance $x$ in $fm$. Our results
are represented by the points. The phenomenological curves
(as derived in \cite{Shuryak_cor}) are given by the two solid lines.
The dashed line
corresponds to the 'vacuum dominance' estimate (see text).
Two dash-dotted lines show the fit described in the text.

\vskip 0.5 cm
\noindent
Fig. 2. Correlators for vector channels with quantum numbers of
$\rho$, $K^*$ and $\phi$ mesons normalized to the free
massless quark correlator in the same channel
plotted versus the distance $x$, in $fm$.
Our results shown by squares, triangles and hexagons, respectively,
should be compared with phenomenological expectations
\cite{Shuryak_cor} (solid curves).
The dashed line corresponds to the 'vacuum dominance' estimate.

\vskip 0.5 cm
\noindent
Fig. 3. Correlators for axial channels normalized to the free
massless quark correlator in the same channel
plotted versus the distance $x$ in $fm$.
 Squares, triangles and hexagons show our results
for the $\bar u d$, $\bar u s$ and $\bar s s$ channels, in this order.
The first one should be compared with the phenomenological expectation
 given by the region in between the two
solid curves \cite{Shuryak_cor}. The 'vacuum dominance' estimate
is represented by the dashed curve.
Three dash-dotted lines show the fits described in the text.

\vskip 0.5 cm
\noindent
Fig. 4. Correlators for isovector scalar channels normalized to the free
massless quark correlator in the same channel
plotted versus the distance $x$ in $fm$.
Squares and triangles  show our results
for  the $\bar u d$ and $\bar u s$ currents, respectively.
The 'lower limit'  to the phenomenological expectation (see text) is
shown by the solid curve, while the dashed one
corresponds to the 'vacuum dominance' estimate.
The dash-dotted lines just guide the eye.

\vskip 0.5 cm
\noindent
Fig. 5. Correlators of the tensor current $j_{\mu\nu}$
normalized to the free {\it pseudoscalar}
massless quark correlator plotted versus the distance $x$ in $fm$.
 Squares, triangles and hexagons show our results
for the $\bar u d$, $\bar u s$ and $ \bar s s$ channels, respectively.
The 'vacuum dominance' estimate
is represented by the dashed curve, and the
dash-dotted curve  shows the fit described in the text.

\vskip 0.5 cm
\noindent
Fig. 6. The vector-tensor (a) and the pseudoscalar-axial (b)
correlators in units of the {\it pseudoscalar} free quark correlator
plotted versus the distance $x$ in $fm$.
The parameters of the solid curve in (a) are those obtained from QCD sum rules
 \cite{Balitsky_Kolesnichenko_Yung}.
The dash-dotted line in (b) shows the fit described in the text.
\vskip 0.5 cm
\noindent
Fig. 7. Correlators for the isoscalar pseudoscalar channels $\eta$ and $\eta'$
 normalized to free {\it pseudoscalar}
massless quark propagator plotted versus distance $x$ in $fm$.
In figure (a) we show the correlators of the
$SU(3)$ singlet and octet currents (open and closed squares), while (b)
represents the singlet-octet mixed correlator.
Phenomenological expectations (see [1])
are shown by the solid curves. The dash-dotted curve in (a)
is our fit for $\eta$,
and the dashed one shows the contribution of the $\eta'$ to the singlet
correlator according to [1].

\vskip 0.5 cm
\noindent
Fig. 8. Correlators for the isoscalar scalar $\sigma$ channel
 normalized to the free
massless quark correlator in the same channel plotted versus the
distance $x$ in $fm$. The full correlator is shown in
Figure (a), whereas (b) only has the connected part.
The  solid curve in (a) is the phenomenological
expectation for the disconnected part.
The long and short-dashed curves in (b) show the total
correlator and  the $\sigma-$meson contribution to the fit, respectively.

\vskip 0.5 cm
\noindent
Fig. 9. Comparison between our results for the correlation functions
(triangles)
and the first lattice results \cite{Negele_etal} (squares).
The four channels
considered  correspond to the isovector ($e.g.$ $\bar u d$)
 pseudoscalar, vector,
scalar and axial the cases. The phenomenological result (solid lines)
were derived in \cite{Shuryak_cor}. All correlators are normalized with respect
to the free quark correlator in the same channel and the distance on the
$x-$axis is in $fm$.
\newpage
\addtocounter{page}{-5}


\end{document}